\documentclass[manuscript,screen]{acmart}
\AtBeginDocument{%
  \providecommand\BibTeX{{%
    \normalfont B\kern-0.5em{\scshape i\kern-0.25em b}\kern-0.8em\TeX}}}

\setcopyright{acmlicensed}
\copyrightyear{2018}
\acmYear{2018}
\acmDOI{XXXXXXX.XXXXXXX}

\acmConference[Conference acronym 'XX]{Make sure to enter the correct
  conference title from your rights confirmation emai}{June 03--05,
  2018}{Woodstock, NY}
\acmISBN{978-1-4503-XXXX-X/18/06}

\usepackage{graphicx}
\usepackage{subcaption}
\usepackage{float}
\usepackage{amsmath}
\usepackage{amsthm}
\usepackage{booktabs}
\usepackage{hyperref}
\usepackage{enumitem}
\usepackage{multirow}
\usepackage{tikz}
\usepackage[edges]{forest}
\usepackage{xcolor,colortbl}
\definecolor{highlightgray}{RGB}{224,224,224}

\begin{document}

\title{Privacy and Security Implications of Cloud-Based AI Services : A Survey}

\author{Alka Luqman} 
\email{alka001@ntu.edu.sg}
\affiliation{%
  \institution{Nanyang Technological University}
  \city{Singapore}
  \country{Singapore}
}

\author{Riya Mahesh}
\email{ee21b112@smail.iitm.ac.in}
\affiliation{%
  \institution{Indian Institute of Technology}
  \city{Madras}
  \country{India}
}

\author{Anupam Chattopadhyay}
\email{anupam@ntu.edu.sg}
\affiliation{%
  \institution{Nanyang Technological University}
  \city{Singapore}
  \country{Singapore}
}
  

\begin{abstract}
This paper details the privacy and security landscape in today’s cloud ecosystem and identifies that there is a gap in addressing the risks introduced by machine learning models. As machine learning algorithms continue to evolve and find applications across diverse domains, the need to categorize and quantify privacy and security risks becomes increasingly critical. 
With the emerging trend of AI-as-a-Service (AIaaS), machine learned AI models (or ML models) are deployed on the cloud by model providers and used by model consumers.
We first survey the AIaaS landscape to document the various kinds of liabilities that ML models, especially Deep Neural Networks pose and then introduce a taxonomy to bridge this gap by holistically examining the risks that creators and consumers of ML models are exposed to and their known defences till date. 
Such a structured approach will be beneficial for ML model providers to create robust solutions. Likewise, ML model consumers will find it valuable to evaluate such solutions and understand the implications of their engagement with such services.
The proposed taxonomies provide a foundational basis for solutions in private, secure and robust ML, paving the way for more transparent and resilient AI systems.
\end{abstract}

\maketitle

\section{Introduction}
\noindent 
Machine learning (ML) is becoming increasingly ubiquitous nowadays, forming part of the computation fabric of many businesses. According to a Fortune Business Insights report \footnote{https://www.fortunebusinessinsights.com/machine-learning-market-102226}, the global ML market size is expected to increase from USD 26.03 billion in 2023 to USD 225.91 billion in 2030. Various prominent cloud providers that used to offer infrastructure as a service (IaaS), platform as a service (PaaS) and software as a service (SaaS) now advertise Machine Learning as a service (MLaaS) in their portfolio. The cloud-based delivery model of ML as a service has reduced the barrier to entry in designing applications that use machine learning techniques.
MLaaS offers machine learning algorithms, models, and tools that users can leverage without the need to build and maintain their own machine learning infrastructure. AI as a Service (AIaaS) is a broader offering that encompasses rule-based systems, symbolic reasoning, natural language processing APIs, chatbot builders etc., in addition to ML tools.
\par

\begin{figure}[b]
\centering
  \includegraphics[scale=0.5]{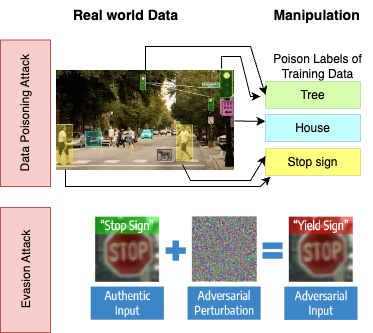}
  \caption{Examples of attacks affecting autonomous vehicles during model training by poisoning input data labels or during inference by manipulating scenes in real-time.}
  \label{fig:attack_eg}
\end{figure}

\subsection{Background}
ML models, especially deep learning models which are becoming increasingly common, are susceptible to various attacks and pose serious safety implications in applications like smart homes, autonomous driving, etc. We show in Figure \ref{fig:attack_eg} how data poisoning attacks on a self-driving vehicle can introduce misleading data into the training dataset, causing the model to learn incorrect patterns. For example, attackers could manipulate the labels of camera images to mislead the model's understanding of the environment. Man-in-the-middle attacks can also exploit vulnerabilities in the model by performing evasion attacks. For instance, adversarial perturbations on road signs may cause the autonomous vehicle to misinterpret critical information, leading to potentially dangerous decisions. Robustness against these attacks is crucial to ensure the safety and reliability of autonomous driving systems, necessitating the development of resilient models and security measures.\par
With the proliferation of custom Large Language Models (LLMs) like ChatGPT and their usage, it is important to assess the privacy risks they present. This recent technology offering has been shown to be vulnerable to prompt injection attacks that expose the training instructions and files used to customize the custom chatbots \cite{yu2023assessing}. Applications leveraging LLMs to read and parse personal emails to act upon their content can pose a serious security concern. Adversaries can craft emails containing specific prompts, forcing the LLM to execute malicious actions, like unauthorized purchases or content summarization with potentially harmful consequences. \par

\subsection{Related Surveys}
There has been a lot of research \cite{10179281, 9734024, 10.1145/3523273} focused on understanding the privacy and security risks associated with ML models. Various surveys have shed light on the vulnerabilities, potential attacks, and privacy concerns in the deployment of ML models. A taxonomy for privacy in ML proposed in \cite{10.1145/3436755} decouples using ML to perform targeted attacks and using ML as an aid to enhance user privacy. Attacks and their impact are judged  in \cite{9294026} based on the time of attack during its lifecycle. \par

These works have contributed significantly to our understanding of the subject. However, as the field of ML advances, the threat landscape evolves, demanding continuous exploration of novel challenges and innovative solutions. To address the ever-changing landscape of ML model security, there is a need for fresh perspectives that can provide insights into emerging threats and privacy-preserving techniques to combat them.  \par
\subsection{Motivation}
\noindent
The motivation of this paper is to highlight the importance of including privacy risks posed by ML models while promising data privacy guarantees in cloud services. Even if the training data is kept encrypted, the Deep Neural Network (DNN) models trained on such private data are still vulnerable to privacy leakage \cite{shokri_mi_2017, hitaj_deepmodelsundergan_2017, zhibo_privacyleakageFL_2019}. It is no longer enough to talk about data protection in silos, but to view it holistically with the ML models that make use of that data. ML attacks targeting edge IoT systems fall outside the scope of this paper. Here we focus on ML solutions hosted in the cloud via AIaaS. \par
\subsection{Contribution}
\noindent
We propose a taxonomy to comprehensively view ML models and its training data based on resource ownership and usage rights. We catalogue the privacy guarantees assured by major cloud service providers like Microsoft, Google, Amazon, Huawei, Alibaba, IBM and Oracle, and map it to the proposed taxonomy. We include common attacks and their defence strategies presented in both theoretical settings as well as open source and private solutions. \par
The rest of the paper is structured as follows. Section II introduces the objectives commonly aspired for in information security and the many ways to achieve them - using protocols or specialized hardware. It then lays out the techniques used in the current cloud infrastructure for ML as a service to guarantee information security. Section III talks about the gaps in ML model security by listing out recent attacks. Section IV proposes the taxonomy considering ownership and usage rights and Section V contains the taxonomy considering defence techniques. \par 


\section{Privacy Guarantees and Threats in the current cloud ecosystem}
\label{sec:current_ecosystem}
\subsection{Security Objectives}
\begin{figure}[h]
\centering
  \includegraphics[scale=0.5]{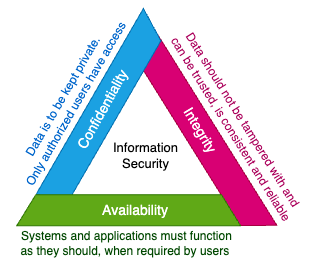}
  \caption{CIA triad for Information Security.}
  \label{fig:cia_data}
\end{figure}

\noindent
Information security encompasses a broad range of practices to protect sensitive information and prevent illegal access and modification of digital information. Many strategies are put forward to ensure the primary security measures of Confidentiality, Integrity and Availability, called the CIA triad. The CIA triad, depicted in Figure \ref{fig:cia_data}, serves as a foundational framework for designing and evaluating security measures and controls, along with other measures like non-repudiation and authentication.\par

\textbf{Confidentiality} is to prevent illegal access to the content of messages, ensuring that private and sensitive information is only accessible to authorized individuals. This is often achieved through measures such as encryption and access controls. Symmetric encryption techniques like DES and AES use a single public key to encrypt and decrypt data, while asymmetric encryption techniques like RSA and ECC use a pair of keys, an openly shared public key for encryption and a private key for decryption. \par

\textbf{Integrity} is to detect and prevent unauthorized modification of messages, ensuring that data and systems are reliable and trustworthy. This is usually achieved using cryptographic hash functions, Message Authentication Codes (MACs) and digital signatures. Hash functions generate unique checksums for data, and obtaining a different checksum indicates that the underlying data has been changed. MACs are generated using a keyed hash function. Digital signatures provide a way to verify the authenticity of data and confirm that it has not been altered by unauthorized parties. \par

\textbf{Availability} is to safeguard against attacks that target the system, ensuring that the platforms through which data is accessed are accessible and operational when needed, without interruption. This involves safeguarding against downtime, denial-of-service (DoS) attacks, and other disruptions. There is no cryptographic measure to prevent this and it is often achieved using backups and redundancy strategies. \par

\textbf{Non-repudiation} is to prevent individuals from denying the validity or origin of actions or transactions. Digital signatures and audit trails help establish this.

\textbf{Authentication} is to verify the identity of users attempting to access information or systems. This is often achieved using public key constructs like digital signatures or private key constructs like Message Authentication Codes.

\subsection{Cryptographic Standards}
\noindent
Cryptographic standards refer to established ciphers for various applications, achieving security objectives mentioned in the previous subsection. Some of the commonly used algorithms are mentioned below.

\noindent
\textbf{Private-Key Cryptography} 
\begin{itemize}
    \item \textbf{AES (Advanced Encryption Standard)} is a symmetric-key algorithm that operates on data blocks of size 128 bits with keys of length 128, 192, or 256 bits. It was developed in a public competition organized by the U.S. National Institute of Standards and Technology (NIST), where the Rijndael algorithm developed by Belgian cryptographers Vincent Rijmen and Joan Daemen, was selected as the winner. AES encryption involves multiple rounds of transformations on the data block, including substitution, permutation, and mixing operations. These rounds are performed on the data using a round key derived from the original encryption key. AES, with its larger key size, is considered more secure than its predecessor DES. It is usually implemented using block ciphers for encryption. The two popular cipher modes are \textit{CBC (Cipher Block Chaining mode)} and \textit{GCM (Galois/Counter mode)}. CBC encrypts data block using XOR chaining while GCM operates with counter mode of encryption and Galois mode for authentication leading to a higher throughput.\par
    \item Lightweight block cipher algorithms like PRESENT, SIMON and SPECK are designed to provide secure encryption with a focus on simplicity and efficiency in resource-constrained environments. \textbf{PRESENT} is known for its small hardware footprint and is used in low-power devices like RFIDs and sensors. It uses a compact and efficient round structure with a minimal number of operations, making it well-suited for resource-constrained devices. \textbf{SIMON} is designed to offer strong security while being highly efficient in terms of both hardware and software implementations. It can be used in a wide range of applications, including lightweight cryptography for small IoT devices, secure communication protocols, and encryption in constrained computing environments. \textbf{SPECK} is a family of lightweight block ciphers designed to provide efficient and secure encryption for various applications such as IoT devices, mobile communication, and constrained computing platforms. \par
\end{itemize}

\noindent
\textbf{Public-Key Cryptography} 
\begin{itemize}
    \item \textbf{RSA (Rivest-Shamir-Adleman)} is an asymmetric-key algorithm which uses a public key for encryption and a private key for decryption. When a user wants to send an encrypted message to a recipient, they use the recipient's public key to encrypt the message. This ensures that only the recipient, who possesses the corresponding private key, can decrypt and read the message. The operations in RSA involve modular exponentiation with large prime numbers, making it computationally infeasible to derive the private key from the public key.\par
    \item \textbf{ECC (Elliptic Curve Cryptography)} is a type of public-key cryptography that is based on the algebraic structure of elliptic curves over finite fields. The strength of ECC lies in the difficulty of calculating the private key from the public key, known as the elliptic curve discrete logarithm problem. It is widely used for securing digital communications, data encryption, and digital signatures due to its strong security properties and efficiency. ECC keys are much shorter than equivalent RSA keys, giving it an advantage in resource-constrained environments.\par
    \item \textbf{PQC (Post-Quantum Cryptography)} refers to the field of cryptography that focuses on developing cryptographic algorithms and protocols that remain secure against attacks from quantum computers. Quantum computers have the potential to break many of the widely used encryption schemes like RSA and ECC, by leveraging quantum algorithms like Shor's algorithm for factoring large numbers and Grover's algorithm for searching unsorted databases.\par
\end{itemize}

\noindent
\textbf{Message Authentication Code and Hash Functions}
\begin{itemize}
    \item \textbf{MAC (Message Authentication Code)} is a cryptographic technique used to verify the authenticity and integrity of a transmitted message. It is derived from the message and a secret key, and is sent alongside the message to ensure that it has not been tampered with during transmission. It utilizes a secret key, which is private to the sender and the recipient, to both generate and verify the MAC. \par
    \item \textbf{MD5 (Message Digest) Algorithm} is a cryptographic hash algorithm that generates a hash code of fixed length (128 bits) for any arbitrary input message. While still in use, the algorithm has shown vulnerability to collision attacks and is found to be cryptographically broken. This has resulted in the adoption of alternative algorithms like SHA for hashing. \par
    \item \textbf{SHA (Secure Hash Algorithm)} converts any input into a condensed hash value of fixed length called as the Message Digest. It is based on Merkle-Damgard construction technique. \textit{SHA-0}, \textit{SHA-1}, \textit{SHA-224}, \textit{SHA-256}, \textit{SHA-384}, \textit{SHA-512} are the existing versions of this family which differ in their strength of security and size of the block that they operate upon.  \par
\end{itemize}

\subsection{Communication Protocols}
\noindent
Cryptographic communication protocols are sets of rules and procedures that govern secure communication between two or more parties over a network or any communication channel. These protocols employ cryptographic techniques to ensure the objectives of confidentiality, integrity, and authenticity of the data being exchanged. Some examples are given below, applicable from lowest to highest layers of the OSI stack. \par
\begin{itemize}
    \item \textbf{SSL (Secure Sockets Layer)} is a cryptographic protocol designed to provide secure communication over a computer network. It was first developed by Netscape Communications in the 1990s to add the HTTPS protocol to their web browser Netscape Navigator. SSL was designed to encrypt data transmitted between a client and a server, ensuring that the data exchanged during these sessions could not be easily intercepted or deciphered by malicious actors. Due to various security vulnerabilities, SSL has been deprecated in favour of TLS.\par
    \item \textbf{MACsec} is a data-link layer encryption method using the IEEE 802.1AE MAC Security Standards. Packets are encrypted and decrypted on the devices before being sent, preventing snooping or wiretapping attacks. This technology is integrated on the network hardware, and so provides encryption on the network hardware with no measurable increase in link latency. \par
    \item \textbf{TLS (Transport Layer Security)} is a cryptographic protocol designed to secure communications over a computer network. It is commonly used to protect data transmitted over the internet, ensuring the privacy, integrity, and authenticity of the data exchanged between two systems. It is commonly used in securing web browsing (HTTPS), email communication (SMTPS, IMAPS), VPN connections etc. TLS 1.3, defined in August 2018, is the most recent version.\par
    \item \textbf{SSH (Secure Shell)} is a cryptographic network protocol used for securely connecting to and managing network devices and servers over an unsecured network. SSH was first designed in 1995 and uses public-key cryptography to allow users to authenticate themselves to a remote computer. It also includes utilities like SCP (Secure Copy Protocol) and SFTP (SSH File Transfer Protocol) for secure file transfers between systems, with encryption and data integrity checks. OpenSSH is the most widely implemented variant.\par
    \item \textbf{S/MIME (Secure/ Multipurpose Internet Mail Extensions)} is a standard for securing email messages with encryption and digital signatures. It is commonly employed in email clients and servers to protect the confidentiality of email content (via encryption) and to verify the authenticity of the sender along with non-repudiation of origin (using digital signatures).\par
    \item \textbf{VPN (Virtual Private Network)} is a protocol that provides secure communication over the internet, allowing users to transmit data privately and securely. It is often used to create secure, encrypted tunnels between clients and servers. It is widely used by organizations to ensure confidentiality of the data transmitted over the network. \par
\end{itemize}

\subsection{Trusted Computing}
\begin{itemize}
    \item \textbf{HSM (Hardware Security Modules)} are robust computing devices that provide cryptographic security to data by performing encryption, decryption and key management processes. They are embedded on secure cryptoprocessor chips and assist in creating digital signatures and certificates.  \par
    \item \textbf{Trusted Execution Environment} \cite{priebe2018enclavedb} is a secure enclave isolated from all system software or hardware outside it. It provides a trusted platform to carry out computational tasks and protects the integrity of model and data from an untrusted host environment during execution. TEEs play a crucial role in providing both server-side and client-side protection\cite{mo2022sok}. The former focuses on preventing threats during model training and inference stages while the latter involves securing the model and client data from vulnerabilities once deployed for performing tasks. \par
    \item \textbf{Homomorphic Encryption} \cite{10.1145/3214303} is a cryptographic technique that allows computation to be performed on encrypted data without having to decrypt it. Mathematical operations like addition, multiplication etc. can be performed on the ciphertext directly through specialized encryption algorithms that maintain certain mathematical relationships between plaintexts and ciphertexts. One popular homomorphic encryption-based MPC protocol is the BFV protocol, which allows parties to securely compute a variety of functions on their private inputs.\par
    \item \textbf{Secure Multi-party Computation} is a technique whereby multiple parties collaborate to compute results on their private data, without revealing each part. This can be used to perform privacy-preserving data analysis, privacy-preserving machine learning, enabling smart contracts in blockchain, etc. MPC protocols are based on Secret sharing, Garbled circuits, Threshold cryptography-based MPC etc.\par
    \item \textbf{Searchable Symmetric Encryption} \cite{10.1145/1180405.1180417} is a cryptographic technique that allows search on encrypted records in databases. The advantage of SSE is in the ability to selectively retrieve parts of documents. SSE is well-suited for scenarios where users need to search within a cloud-hosted encrypted database without revealing the plaintext data. Most SSE schemes today utilize secure indexes generated from pre-processed document-keyword pairs while encrypting documents using a symmetric encryption algorithm. The server compares query tokens with messages in the index and matched documents are returned to the data user.\par
\end{itemize}

\subsection{Current privacy guarantees}
\begin{figure}[t]
\centering
  \includegraphics[scale=0.5]{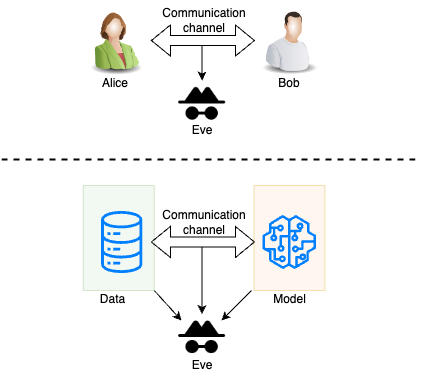}
  \caption{Extending the threat model of cryptography systems to the ML ecosystem.}
  \label{fig:alice_bob}
\end{figure}

\noindent 
Data outsourcing is a widely practised cloud service where users and companies enlist third parties to store and manage their data. This is favourable in the face of high costs of storage and of skilled professionals to design and manage secure in-house storage systems. But this comes at the cost of potentially exposing private data \cite{10.1145/1755688.1755690}.
Encryption is what enables data privacy on the Internet today. If an entity Alice is sending a private message to Bob, as shown in Figure \ref{fig:alice_bob}, the message is encrypted and the only parties with access to the key are Alice and Bob. The private data cannot be understood by any third party who tries to intercept the message while it is in transit through a communication channel. \par

Existing cloud providers provide data encryption at storage nodes (Data at rest), communication links (Data in transit) and compute nodes (Data in use). This categorization is according to the 3 states of digital data. 
\begin{enumerate}
    \item \textbf{Data at rest} refers to data that is stored on physical or virtual storage devices. Data encryption at rest involves encrypting this data to protect it from unauthorized access if the physical storage media is compromised or stolen. It requires customers to use effective key management systems. The keys are stored securely and rotated regularly to minimize any risk of exposure. 
    \item \textbf{Data in transit} refers to data that is being transmitted between devices or services within a cloud environment or between the cloud and external systems. Data is encrypted before it leaves the source and decrypted upon arrival at the destination to maintain confidentiality. Common protocols for encrypting data in transit include HTTPS for web traffic, SSL/TLS for secure communication, and VPNs for secure network connections.
    \item \textbf{Data in use} refers to data that is actively being processed by applications or services in the cloud. This data resides in computer RAM, CPU caches, registers etc. Cryptographic protocols like homomorphic encryption and secure multi-party computation perform computation operations on encrypted data. Another alternative is the use of secure enclaves, which maintains the data in encrypted form in RAM, but in clear text inside CPU caches and registers.
\end{enumerate}

\begin{table*}[ht!]
\caption{Data privacy guarantee in current cloud ecosystem.}
\label{tab:data_privacy}
\vspace*{-1em}    
\resizebox{\columnwidth}{!}{%
\begin{tabular}{p{2cm}p{5cm}p{5cm}p{4.5cm}}
 \toprule
 \textbf{Cloud provider} & \textbf{At rest} & \textbf{In transit} & \textbf{In use}\\
 \midrule
 
 \multirow{5}{2cm}{MS Azure \cite{microsoftProtectionCustomer, microsoftTechnicalReference, microsoftDoubleEncryption}} & AES-256 in CBC mode &  VPN (utilizing IPsec/IKE encryption) & Confidential computing \\
 & FIPS 140-2 compliant & TLS 1.2 \& 1.3  & (Intel SGX) \\
 & & SSH (on Linux VMs)&  \\
 & & SMB 3.0 (between Azure virtual networks)& \\
 & & IPSec (between on-prem network and Azure network)&  \\
 & & MACsec (LAN between datacentres)&  \\
\hline

 \multirow{2}{2cm}{AWS \cite{amazonDataPrivacy, amazonEncryptingFile}} & AES-256 & TLS 1.2, AES-256 & Confidential computing \\
 & & AWS Key Management Service, hardware based HSM & (Nitro enclave) \\
\hline

  \multirow{5}{2cm}{GCP \cite{googleGoogleSecurity}} & Disks created after 2015 use AES-256 (using library Tink) & TLS (default) uses Google's implementation called BoringSSL & Confidential computing (AMD Secure Encrypted Virtualization) \\
 & Disks created before 2015 use AES-128 & IPSec tunnels & \\
 & FIPS 140-2 compliant & S/MIME for Gmail &  \\
 & & Managed SSL certificates & \\
 & & AES-128 (ALTS) for Layer 7 encryption &  \\
 \hline
 \multirow{3}{2cm}{Huawei \cite{huaweiSecurity, huaweiDataSecurity}}&AES-256& TLS 1.2, SSL, VPN & Trusted Intelligence Computing Service \\
&Key Mangement Service&& (ARM Trusted Firmware v1.3)\\
\hline
\multirow{5}{2cm}{IBM \cite{ibmEncryptionData}} & AES - 256 in CBC mode&TLS 1.3&Confidential Compute \cite{ibmWhatConfidential}\\
&RC4-128 encryption with MD5-128 &AES in CBS or GCM mode& (Intel SGX)\\
&AES-128 encryption with MD5-128 &RSA, ECDSA, ECDHE algorithms & Fortanix Runtime Encryption\\
&AES-256 encryption with SHA-256 && Platform\\
\hline
\multirow{4}{2cm}{Oracle \cite{oracleCloudSecurity}}&AES 256 SHA-256 with RSA& TLS 1.1 \& 1.2 &Confidential computing \\
&TDES (Triple Data Encryption Standard) 168 bits&&(AMD Infinity Guard)\\
&Regional Encryption algorithms : ARIA, SEED and GOST &&\\
\hline
\multirow{3}{2cm}{Alibaba \cite{alibabaSecurity}} & HSM, AES-256 &SSL, TLS 1.2 \& 1.3&Intel SGX enclaves\\
&KMS at 2 levels Customer Master Key and Data Encryption & VPN Gateway &\\
&Key &Alibaba Cloud Smart Access Gateway&\\
\bottomrule
\end{tabular}
}
\vspace*{1em}      

\caption{Model privacy guarantee in current cloud ecosystem.}
\label{tab:model_privacy}
\vspace*{-1em} 
\resizebox{\columnwidth}{!}{%
\begin{tabular}{p{2cm}p{5cm}p{5cm}p{4.5cm}}
 \toprule
 \textbf{Cloud provider} & \textbf{At rest} & \textbf{In transit} & \textbf{In use}\\
 \midrule
 MS Azure \cite{microsoftTechnicalReference} & Encrypt model using Azure managed encryption key & Encrypt network traffic using TLS 1.2 & Confidential computing \\
 AWS \cite{amazonDataProtection} & Encrypt model using AWS managed encryption key & Encrypt network traffic using TLS 1.2 & Confidential computing \\
 GCP \cite{googleCustomermanagedEncryption} & Encrypt model using Customer Managed Key & Encrypt network traffic using TLS & Confidential computing \\
 Huawei \cite{huaweiSecurity} & Encrypt model using Customer Managed Key & Encrypt network traffic using TLS 1.2 & Trusted intelligence computing service \\
 IBM \cite{ibmWhatConfidential} & Encrypt model using Customer Managed Key & Encrypt network traffic using TLS 1.3 & Confidential computing \\
 Oracle \cite{oracleCloudSecurity} & Encrypt model using Customer Managed Key & Encrypt network traffic using TLS & Confidential computing \\
 Alibaba \cite{alibabaSecurity} & Encrypt model using 2 Customer Managed Keys & Encrypt network traffic using TLS & Confidential computing \\
 \bottomrule
\end{tabular}
}
\end{table*}

The underlying technologies at each step are consolidated in Tables \ref{tab:data_privacy} and \ref{tab:model_privacy} to see how privacy is currently guaranteed in some major cloud providers, for data and ML models respectively. Confidential computing is overwhelmingly used to address attacks on data during processing by isolating it exclusively to the CPU handling the processing. The information in the tables also emphasize how the guarantees for model privacy currently are piggybacking on existing definitions of data privacy. In light of the widespread adoption of ML applications, these distinctions are no longer enough to encompass all possible interplays of ML model usage with the privacy of its underlying training data. There is a wide range of attacks on ML models which expose additional attack surfaces that a malicious actor may exploit to compromise the privacy of personal and sensitive information. The training data becomes an additional attack vector that threatens the robustness of the ML model. \par
The artefacts in a cloud based ML solution include 
\begin{enumerate}
    \item Training dataset - This includes the private images or textual information used to train an ML model.
    \item Trained ML model - This can be the structure of the tree and the decision criteria in case of a decision tree, the weights of a DNN, or a reward matrix in RL. 
    \item Test dataset - It is used to statically evaluate the model performance.
    \item Inference data - This is the private data of consumers who use the service.
\end{enumerate} 
In the subsequent discussions of the taxonomy we use the term \textit{DATA} to refer to the training dataset, test dataset and inference data, while \textit{MODEL} refers to the trained ML model. An \textit{ML model consumer} is an entity or system that utilizes the model for predictions or decision-making and \textit{ML model provider} is the entity responsible for creating, maintaining, and supplying the ML model and its related services to the consumers. Table \ref{tab:all_attacks} categorizes the numerous attacks on data and model which have been proposed in the recent past. The attacks mentioned in this paper are a representative selection of the works till date and not an exhaustive list.\par

\begin{table*}[ht!]
\caption{Attacks on Data and Model}
\resizebox{\columnwidth}{!}{%
\begin{tabular}{p{1cm}p{3cm}p{6cm}p{4.5cm}p{1.5cm}}
\toprule
\textbf{Attack type} & \textbf{Attack name} & \textbf{Description} & \textbf{Examples} & \textbf{Access Type}\\
\midrule
\multirow{15}*{\rotatebox[origin=c]{90}{\textbf{Exploratory}}}&Adversarial Transfer Attack& Attacker trains a substitute model to study and transfer the attacks on this model to the target model&Transfer Attack\cite{cheng2019improving} &Blackbox\\
\cline{2-5}
&Membership Inference Attack & Adversary determines if a particular data point was a member of the training dataset&Shadow Attack\cite{cretu2023re}&Whitebox\\
\cline{4-5}
&&&GAN based Membership Inference attack\cite{jia2019memguard}&Blackbox\\
\cline{2-5}
&\multirow{4}*{Model Inversion Attack} & Attacker aims to reconstruct the training data by accessing the model&Activation Maximization \cite{nanfack2023adversarial} &Whitebox\\
\cline{4-5}
&&&GAN and Reinforcement Learning based Model Inversion\cite{han2023reinforcement}&Blackbox\\
\cline{2-5}
&Model Reconstruction Attack& Adversary reconstructs the trained model &Local Model Reconstruction Attack\cite{driouich2022local}&Whitebox\\
\cline{2-5}
&Attribute Inference Attack& Adversary reconstructs the model attributes from its local update stages & Attribute Reconstruction Attack \cite{lyu2021novel} & \\
&&&Model Inversion Attribute Inference Attack \cite{mehnaz2020black} & Blackbox \\
\cline{2-5}
&Label Inference Attack&Attacker aims to infer the label of the data points&Label Inference Attack against Ver-
tical Federated Learning\cite{fu2022label}&\\
&&&Label Only Membership Inference Attack\cite{choquette2021label}&Blackbox\\
\cline{2-5}
&Infer Class Representatives& Attacker infers the features of every class to construct representative samples for each of them&Generative Adversarial Network-based infer class representative attack\cite{wang2019beyond}&Blackbox\\
\midrule
\multirow{6}*{\rotatebox[origin=c]{90}{\textbf{Evasion}}}&\multirow{4}*{Model Evasion} & Crafting adversarial examples to introduce perturbations to the input data based on optimization techniques to misclassify them  &Carlini and Wagner attack\cite{lin2021ml}& \multirow{4}*{Whitebox}\\  
&&&Zeroth Order Optimization\cite{lin2021ml}&\\
&&&Deepfool\cite{lin2021ml}&\\
&&&Fast Gradient Sign Method\cite{lin2021ml}&\\
\cline{3-5}
&& The attacker uses trained Generative Adversarial Network to introduce adversarial features to evade the model & GAN Attack against intrusion detection \cite{lin2022idsgan} & Blackbox \\
\midrule
\multirow{5}*{\rotatebox[origin=c]{90}{\textbf{Poisoning}}}&\multirow{5}*{Data Poisoning}& Injecting adversarial action choices into the agent's trajectory &Action poisoning Attack\cite{liu2021provably}&\multirow{4}*{Blackbox}\\
\cline{3-4}
&&Attacker crafts inputs with similar feature representation provoking the model to misinterpret the data &\label{1}Feature Collision Attack\cite{zhu2019transferable}&\\
\cline{3-4}
&&Adversary mislabels the input dataset&Label Flipping Attack\cite{shahid2022label}&\\
\cline{3-5}

&&Attacker attaches a patch to the image that leads to misclassification of the data into an incorrect class &Adversarial Patch Attack\cite{brown2017aurko}&Whitebox\\
\cline{2-5}
&\multirow{3}*{Model Poisoning}&Poisoning local model updates during the training phase&Whitebox online
model poisoning attack on Production Federated Learning\cite{shejwalkar2022back}&Whitebox\\
\cline{3-5}
&&Adversary poisons the model architecture or data during training phase with hidden triggers which are exploited during the deployment stage to misclassify data&Backdoor Attack on CNN  \cite{dumford2020backdooring}&Whitebox\\
\bottomrule
\end{tabular}
\label{tab:all_attacks}
}
\end{table*}

\subsection{Threat model in the Cloud}
A knowledge base of real-world attacks is provided by MITRE ATT\&CK, to be used by organizations who want to develop effective cybersecurity practices. They maintain a continuously evolving list of attack techniques \footnote{https://attack.mitre.org/matrices/enterprise/cloud/} which grants an attacker different levels of access to a cloud enterprise solution that provides IaaS or SaaS. Attacks are grouped by reason of attack a.k.a tactics, i.e. the goal of the adversary during each action. The different goals of an adversary trying to attack at various stages of an enterprise solution is listed in Table \ref{tab:mitre}. The matrix also provides different tried and tested approaches for the attack's detection and mitigation. \par

\begin{table}[h!]
\caption{Tactics listed in the ATT\&CK matrix curated by MITRE for Cloud Enterprise. \colorbox{highlightgray}{Highlighted rows} represent tactics which are also applicable to modern ML solutions.}
\begin{tabular}{p{3cm}p{10cm}}
 \toprule
 \textbf{Tactic} & \textbf{Explanation}\\
 \midrule
 Initial Access & Techniques used to gain an initial foothold within a network.\\
 Execution & Techniques by which the adversary runs malicious code on a local or remote system. \\
 \rowcolor{highlightgray} Persistence & Tactics that the adversary uses to maintain their foothold across any interruptions like system restarts or credential changes.\\
 Privilege Escalation & Tactics by which the adversary tries to gain elevated permissions on a system or network like root level or administrator access.\\
 Defense Evasion & Methods that the adversary uses to evade detection while the system is compromised.\\
 Credential Access & Methods used by the adversary to steal user login credentials.\\
 Discovery & Techniques used by the adversary to explore and gain knowledge about the system or network they are attacking.\\
 Lateral Movement & Techniques the adversary uses to move through the system once it has entered, to reach the target of the attack.\\
 \rowcolor{highlightgray} Collection & Techniques used by the adversary to gather information about the system which will aid in the attack.\\
 \rowcolor{highlightgray} Exfiltration & Techniques to steal the data.\\
 \rowcolor{highlightgray} Impact & Techniques to disrupt the availability of a system or destroy its data.\\
 \bottomrule
\end{tabular}
\label{tab:mitre}
\end{table}

Certain tactics in this categorization also apply to ML solutions. In the \textit{persistence} tactic, an adversary maintains prolonged access to the system. This is manifested in Backdoor ML attacks, where malicious actors manipulate an ML model during its training or deployment to introduce a hidden behavior, similar to a backdoor or trojan. This backdoor behavior is designed to trigger specific responses or outputs when the model encounters specific, often subtle, input patterns or trigger conditions. Such attacks can compromise the integrity and security of the ML system. Attacks to infer the presence of certain samples in the data used to train an ML model, including its statistical properties demonstrate the tactics of \textit{collection} and \textit{exfiltration}. Evasion attacks in ML refer to the process of crafting perturbed input data, called adversarial examples \cite{szegedy2013intriguing, 10.1145/3243734.3264418}, to misclassify it during inference or testing. Such adversarial techniques along with attacks that manipulate the model inference at run time exhibit the \textit{impact} tactic. We review these attacks in detail in the next section, after a brief primer on ML techniques. \par

\section{Threats in the ML ecosystem}
\label{sec:current_threats}
\subsection{Machine Learning Overview}
\noindent
Machine Learning (ML) has witnessed remarkable growth in the past decade, transforming how we interact with technology, handle data, and solve complex problems. There are three fundamental paradigms in \textbf{Classical ML}, based on the nature of the feedback available to the learning system.
\begin{itemize}
    \item \textbf{Supervised Learning} trains a model on labeled data with explicit input-output pairs. Example use cases include image classification, stock price prediction via regression etc.
    \item \textbf{Unsupervised Learning} discovers patterns or structures in data without explicit labels. Some applications that use this include customer segmentation via clustering, fraud detection and generative modeling.
    \item \textbf{Reinforcement Learning} trains intelligent agents to make a sequence of decisions in an environment in order to achieve a goal or maximize a cumulative reward. It involves exploring the environment through interaction with it and exploiting the information present in it. It is commonly used in urban traffic management, mastering motor control in robotics and playing strategy games like Go, Poker, Dota2, etc. \par
\end{itemize}

\noindent
\textbf{Deep Neural Networks (DNNs)} introduced the concept of neural networks with multiple hidden layers. Convolutional Neural Networks (CNNs) are well suited for working with image data on tasks like image recognition and image segmentation. Recurrent Neural Networks (RNNs) perform well with sequential data, and are used in speech recognition and language translation. \textbf{Transformer Networks} use attention mechanisms \cite{NIPS2017_3f5ee243} to process sequences of data in parallel, enabling unprecedented improvements in natural language processing and machine translation. They are also used in computer vision and reinforcement learning.\par

\textbf{Transfer Learning} \cite{9134370} is a technique that extends the knowledge learned on one domain to a different but related task in another similar domain. Transfer learning is especially valuable when there is limited data to train a target task, as it leverages the knowledge acquired from a larger or more relevant source task. Models like BERT highlight the power of pre-training and fine-tuning required for such tasks.\par

\textbf{Federated Learning} \cite{9084352}is a collaborative learning approach that allows multiple devices or entities to collaboratively train a shared model while keeping their data decentralized and private. Instead of sending raw data to a central server for training, federated learning reduces privacy concerns by training models locally on each device and sending the local model parameters to a central server for aggregation. \par

\textbf{Split learning} \cite{khan2023split}is another privacy-preserving collaborative learning approach that shares activation maps between devices after using Homomorphic Encryption to encrypt them. Activation maps represent the activation of neurons in a neural network generated during forward propagation in response to input data. They can provide insight into the features learnt at different layers of the neural network. \par

\begin{figure}[h!]
\centering
  \includegraphics[scale=0.32]{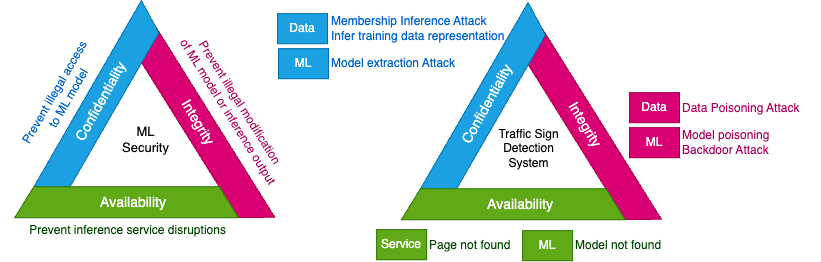}
  \caption{Extending the CIA triad for Model Security and using it to depict the security attacks possible on an example ML application that identifies physical traffic signs on roads to control driving in autonomous vehicles.}
  \label{fig:cia_model}
\end{figure}

\subsection{Privacy Attacks} 
\noindent
Privacy Attacks refer to practices that aim to infer private information of participants, be it individuals or organizations. While data at rest is vulnerable to theft by malicious agents, deep learning models are also vulnerable to reveal their underlying training data and its characteristics due to deep leakage via gradients. In these attacks the private data used to train ML models and the model itself are the assets under attack. \par

\textbf{Data privacy attacks} refer to techniques where the adversary tries to leak private and sensitive information from the dataset used to train an ML model. Some examples of such attacks include -
\begin{itemize}
    \item Inference of class representatives - Such attacks deduce the features of importance to different classes in the training data and generate samples in real time, which appear to come from the same distribution as the training data. \par
    \item Inference of statistical properties - These attacks attempt to infer satistical features like when did a particular sample first appear in the training set, the ratio of men to women etc.\par
    \item Inference of sensitive attributes - In these attacks the adversary leverages additional public datasets to re-identify individuals or their private information used to train an ML model. Given partial information from a patient's medical record, \cite{184489} attempts to infer each individual's genotype information.  \par
    \item Reconstruction of training input and labels - These attacks target ML models to reverse engineer or recover sensitive information from the model's training data. The attacker typically has access to the model's predictions and tries to infer the inputs or labels that led to those predictions.
    \item Membership Inference Attacks - These attacks gauge if an input sample was part of the training dataset used to train the ML model. It is usually exposed by non zero gradients in the NN embedding layer, or by overfit or complex models with high number of parameters. This is a Black Box attack, which can succeed without direct access to the ML model, just by listening to communication channel. Mitigating membership inference attacks often involves adding noise to the training data or applying privacy-preserving techniques like differential privacy to protect against data leakage.\par
\end{itemize}

\textbf{Model privacy attacks} are methods by which an attacker tries to steal the ML model itself, in the case of models which are not freely available to the public. These include -
\begin{itemize}
    \item Model inversion attacks try to infer training data using query-response pairs by observing the predicted label and confidence scores of prediction. This is usually a black box attack, but can become a White Box attack, after a successful extraction attack on the model.
    \item Model reconstruction attack is a superset of the model inversion attack. A GAN reconstruction is used to recreate the ML model. This attack is also applicable to non neural network ML models which store feature metadata.
    \item Model extraction is a black box attack that tries to learn the same task as the target with at-par accuracy, by learning where the decision boundaries of the target model lie. It has been shown that hyperparameters can also be successfully extracted.
\end{itemize}

\subsection{Security / Robustness Attacks} 
\noindent
Robustness Attacks aim to compromise the global ML model by manipulating NN layer gradients or poisoning the training data. In these attacks the quality of the ML model is the commodity of value. \par
\textbf{Data robustness attacks} target the training data residing on devices and attempts to modify it so that the ML model does not reach acceptable accuracy levels.
\begin{itemize}
    \item Data poisoning attacks like the Label Flipping attack modify the target class of training data.
\end{itemize}

\textbf{Model robustness attacks} tamper with the model training process to degrade the prediction accuracy of individual or global models.
\begin{itemize}
    \item Model poisoning attacks like the Byzantine attack work by devices uploading malicious gradients so that the averaged model moves further away from the global minima. Sybil attacks are another form of model poisoning where multiple clone devices are spawned and controlled by one powerful device to tamper with the gradient convergence.
    \item Backdoor attacks embed a backdoor task during model training, which gets activated by a specific input. Examples of such attacks include Adversarial Backdoor Embedding, Clean Label feature Collision etc. 
    \item Evasion attacks are white box attacks that add small perturbations to the model input in order to change the predicted label.
\end{itemize}

\subsection{Threat Model for an ML service} 
\begin{figure}[b!]
\centering
  \includegraphics[scale=0.4]{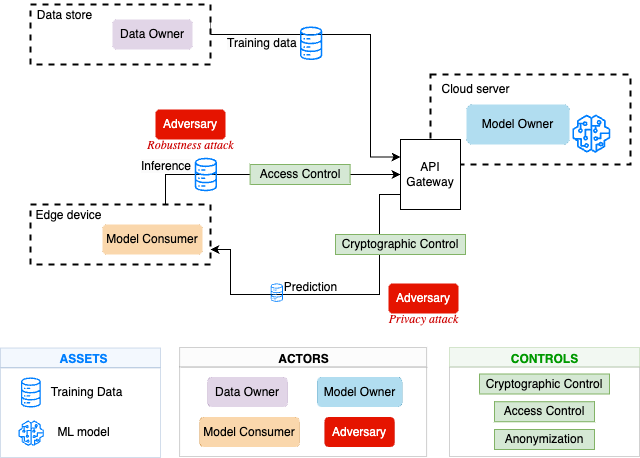}
  \caption{Threat model on a Data Flow Diagram of a simple ML-as-a-service model. The actors interact with the system using assets, which are the resources targeted by an attacker. The dashed lines represent threat boundaries. Controls are placed at vulnerable attack surfaces to mitigate risk.}
  \label{fig:threat_model}
\end{figure}

\noindent
Threat models are used in cybersecurity to systematically identify and mitigate potential risks in a system. Resources which are targeted by an attacker are called assets. In any ML application, the most important assets are the ML model and the data used to train it. Entities that interact with the system or play specific roles are called actors. Here we focus on the data owner who provides training data, the model owner who builds and trains the ML model, and the model consumer who uses this ML service for inference. The other actor is an adversary or a threat agent, which can either be active or passive. A passive adversary is an honest-but-curious participant who tries to leak private information about the underlying training data. An active adversary is a malicious participant who tries to sabotage the ML model's performance. \par
Trust boundaries are drawn to logically separate areas where trust is assumed and areas where trust is not assumed. In the threat model shown in Figure \ref{fig:threat_model} the dashed lines represent trust boundaries among different devices. Controls are countermeasures put in place to mitigate the vulnerabilities identified in the system. Preventions are controls that may completely prevent the possibility of a particular attack, e.g. access controls to authorize certain operations. Mitigations are controls to reduce either the likelihood or the impact of a threat, e.g. anonymization defences like differential privacy, cryptographic controls that encrypt data etc. \par 

\subsection{Threat Model considering the lifecycle of an ML model} 

\begin{figure*}[h!]
\usetikzlibrary{chains,shapes.symbols}
\definecolor{myc}{RGB}{240, 220, 255}
\definecolor{myc1}{RGB}{245, 240, 255}
\definecolor{myc2}{RGB}{213, 232, 212}
\tikzstyle{process} = [rectangle, minimum width=2.5cm, minimum height=0.5cm, text centered, text width = 2.5cm, draw=black]
\begin{subfigure}[t]{\textwidth}
\centering
\resizebox{0.95\textwidth}{!}{%
\begin{tikzpicture}[nodes={shape=signal,signal from=west, signal to=east,align=center,fill=myc,font=\sffamily,on chain,minimum width=2.5cm,text width = 2.5cm, text centered, minimum height=3em,inner xsep=1em},start chain=going right,node distance=1ex]
    \path node(node1)[signal from=nowhere]{Data\\Collection}     node(node2){Data\\Cleaning} 
    node(node3){Data\\Labeling}  node(node4){Feature \\ Engineering}  node(node5){Model Training\\ / Evaluation} node(node6){Model \\ Inference};
    \node (pro1) [process, below of=node1,fill=myc1,xshift=-0.15cm,yshift=-2cm] {\\ Poisoning\\ \cite{brown2017aurko} \\~\\};
    \node (pro2) [process, below of=node2,fill=myc1,xshift=-0.15cm,yshift=-2cm] {\\ Poisoning\\ \cite{koh2022stronger}\\~\\};
    \node (pro3) [process,  below of=node3,fill=myc1,xshift=-0.15cm,yshift=-2cm] {\\ Poisoning\\ \cite{shahid2022label}\\~\\};
    \node (pro4) [process,  below of=node4,fill=myc1,xshift=-0.15cm,yshift=-2cm] {\\ Evasion\\ \cite{lin2022idsgan}\\~\\};
    \node (pro5) [process,  below of=node5,fill=myc1,xshift=-0.15cm,yshift=-2.62cm] {\\ Evasion\\ \cite{lin2021ml}\\~\\Poisoning \cite{liu2021provably} , \cite{dumford2020backdooring}\\~\\};
    \node (pro6) [process, below of=node6 ,fill=myc1,xshift=-0.15cm,yshift=-2.78cm] {\\ Exploratory\\ \cite{cretu2023re},\cite{nanfack2023adversarial},\cite{han2023reinforcement},\\ \cite{driouich2022local},\cite{lyu2021novel},\cite{wang2019beyond}\\~\\ Poisoning\\ \cite{dumford2020backdooring}\\~\\};
    \node (eg1) [process, below of=pro1,fill=myc2,xshift=-0.15cm,yshift=-2cm] {\\ Cryptographic\\ };
    \node (eg2) [process, below of=pro2,fill=myc2,xshift=-0.15cm,yshift=-2cm] {\\ Cryptographic\\};
    \node (eg3) [process,  below of=pro3,fill=myc2,xshift=-0.15cm,yshift=-2cm] {\\ Cryptographic\\};
    \node (eg4) [process,  below of=pro4,fill=myc2,xshift=-0.15cm,yshift=-2cm] {\\ Anonymization\\};
    \node (eg5) [process,  below of=pro5,fill=myc2,xshift=-0.15cm,yshift=-2.62cm] {\\ Anonymization\\Cryptographic\\};
    \node (eg6) [process, below of=pro6 ,fill=myc2,xshift=-0.15cm,yshift=-2.78cm] {\\ Acccess based\\Cryptographic\\};
\end{tikzpicture}
}%
\caption{The first row breaks down the ML lifecycle into stages. The attacks (detailed in Table \ref{tab:all_attacks}) and defences applicable at each stage are shown in the second and third rows respectively.}
\label{fig:ml_lifecycle}
\end{subfigure}
\vspace{2em}

\begin{subfigure}[t]{0.45\textwidth}
\centering
\includegraphics[scale=0.4]{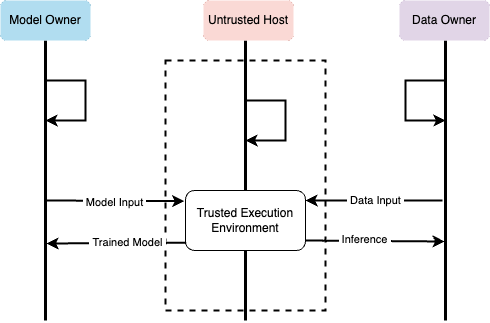}
\caption{During model training}
\label{fig:model_training}
\end{subfigure}
\begin{subfigure}[t]{0.45\textwidth}
\centering
\includegraphics[scale=0.4]{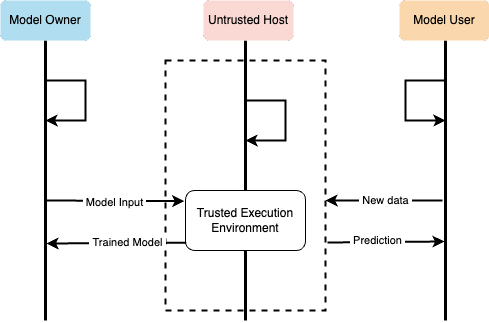}
\caption{During model inference}
\label{fig:model_inference}
\end{subfigure}

\caption{Threat model on different stages during the lifecycle of an ML model is shown in Figure \ref{fig:ml_lifecycle}. The attack surfaces involved during 2 specific stages is further elaborated in Figures \ref{fig:model_training} and \ref{fig:model_inference}.}
\end{figure*}

\noindent 
Figure \ref{fig:ml_lifecycle} depicts a threat model which takes into account the time when an attack occurs in the lifecycle of an ML model. This criterion provides a deeper insight into the precise phases of model design and deployment which exhibit vulnerability. This allows application developers to choose appropriate defence strategies at different stages. \par

The lifetime of an ML model comprises of the following stages:
\begin{itemize}
\item Data Collection - The first step under data preparation involves obtaining data in its raw form. Tampering the dataset at this stage will propagate the error throughout the subsequent phases.
\item Data Cleaning - The data from the previous step is filtered by removing noises and detecting anomalies. Disregarding useful data as unwanted could mislead the model.
\item Data Labeling - Assigning label to data points, usually in the context of supervised machine learning is the next step. Attacks manifest through intentional mislabelling or manipulating the data to provoke the model for erroneous classification.
\item Feature Engineering - It involves selection, transformation  and augmentation of data to extract relevant features and attributes as inputs to the model. Introducing false data that mimics the original features or inducing the model to generate adversarial outputs by strategically biasing specific features are the possible forms of attack vulnerabilities.
\item Model Training and Evaluation - Carrying out attacks during the training and evaluation phase involves manipulating the learning trajectory of the model, modifying gradients and introducing perturbations to inputs within the hidden layers.  
\item Model Inference - Attacks primarily focus on exploiting the model after training to extract information about the model parameters, reconstruct the model and infer the membership and the representative class of a particular data point.
\end{itemize}
The attacks listed in Table \ref{tab:all_attacks} are mapped to these stages where applicable in Figure \ref{fig:ml_lifecycle} and their corresponding defences are also mentioned.
  

\section{Taxonomy considering Ownership and Usage rights}
\label{sec:data_model_taxonomy} 
\noindent
We will now merge the cloud security guarantees, as discussed in Section \ref{sec:current_ecosystem}, with the threats presented Section \ref{sec:current_threats} to provide a new perspective through which to analyze the cloud ecosystem.\par
In today's world of connected devices, the data required to train an ML model may not all be present on one device. Data is generated by multiple autonomous devices in Wireless Sensor Networks \cite{1507522}, which work together in the Internet of Things \cite{ASGHARI2019241} to achieve one or many objectives like traffic prediction \cite{9316934}, learning smart home policies \cite{9097597} etc. With the advent of cloud computing, services like Google's BigTable \cite{10.1145/1365815.1365816} have demonstrated better flexibility and higher performance by distributing storage resources rather than having the data aggregated at a single data centre. \par

Legal restrictions like the General Data Protection Regulation (GDPR) \cite{doi:10.1080/13600834.2019.1573501} regulate how data of people in the European Union is collected, stored and processed. Similar legislations in other countries include the Personal Data Protection Act (PDPA) \cite{CHIK2013554} in Singapore, the Data Protection Act in the United Kingdom \cite{uk_dpa} and the Consumer Privacy Bill of Rights (CPBR) \cite{privacy_gaff} in the United States. These policies deter or complicate the process of aggregating data across devices and regions. \par

\begin{figure*}[h!]
\begin{center}
\begin{subfigure}[t]{0.8\textwidth}
\centering
  \includegraphics[width=\textwidth, scale=0.5]{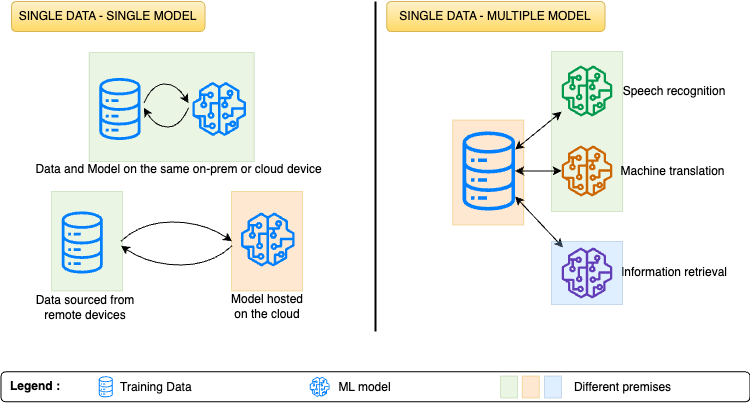}
  \caption{Applications with single data owner}
  \label{fig:data_model_taxonomy_split1}
\end{subfigure}

\vspace{2em}    

\begin{subfigure}[t]{0.8\textwidth}
\centering
  \includegraphics[width=\textwidth, scale=0.5]{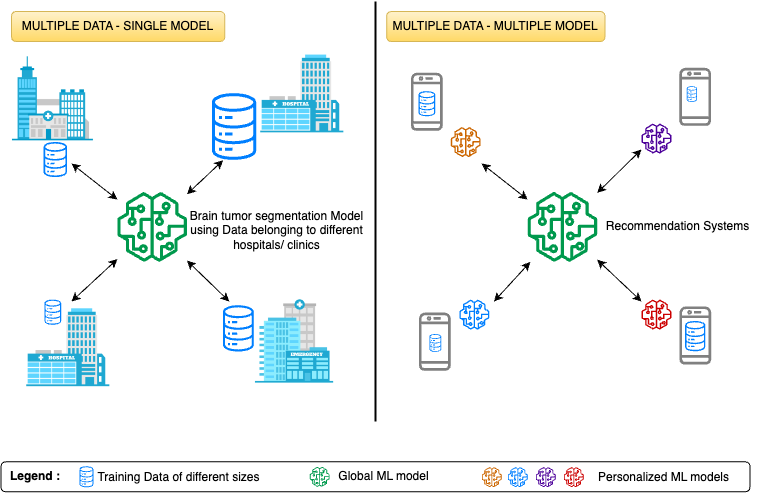}
  \caption{Applications with multiple data owners}
  \label{fig:data_model_taxonomy_split2}
\end{subfigure}
\end{center}

\caption{Proposed taxonomy of Data-Model organization considering ownership and usage rights.}
  \label{fig:data_model_taxonomy}
\end{figure*}

We propose a taxonomy inspired by Flynn's taxonomy which is used to classify computers based on the number of instruction streams and data streams in their architecture. Flynn's taxonomy was introduced in 1966 to classify parallelism in computers and it captures what \textit{instruction} each processor executes along with what \textit{data} it works on. We take a similar view to classify ML applications based on \textit{model objectives} and the distributed nature of the \textit{data} used to train it. Figure \ref{fig:data_model_taxonomy} presents examples for each setting in the proposed taxonomy. \par

\subsection{Single Data Single Model (SDSM)}
Amongst the devices participating in ML, the training data can reside at a single device or be distributed across devices, with varying levels of trust between these devices. The data owner trains a single ML model which is then available for private or public use.\par
This includes applications where one ML model is trained on data present on the same machine or data sourced from a different machine, e.g. applications using third-party ML as a Service. It also includes ML applications that aggregate data from multiple machines that still fall under the purview of a single owner, e.g. inventory management and real-time vehicle tracking for logistics optimization, smart farming, remote sensing and diagnostics. \par

\subsection{Single Data Multiple Model}
A single data owner with access to a large corpus of data can create multiple ML models such that they still retain ownership of the models, but it is available for use by anyone.
This includes Multi-task learning applications where commonalities and differences across tasks are leveraged to learn multiple ML objectives from a shared dataset. \cite{zhang_mtl_2014} use head pose estimation and facial attribute inference as auxiliary tasks for facial landmark detection; \cite{pmlr-v70-arik17a} jointly predict the phoneme duration and frequency profile for text-to-speech conversion. Applications that use transfer learning, like detecting Alzheimer's disease \cite{s19112645} using a modified AlexNet model architecture pre-trained on ImageNet, also fall into this category. \par

\subsection{Multiple Data Single Model}
In situations where a large public corpus of data is not available, we can aggregate the learning from multiple private datasets. Storing and processing large volumes of data using a cluster of machines is well-studied in distributed database systems and distributed computing. Tools like Apache Hadoop and Apache Spark used prevalently in Big Data analytics utilize paradigms like MapReduce to perform statistical aggregation and draw insights from multiple data sources.\par
The numerous devices or clients represent multiple data owners whose datasets may differ in volume and quality, and cannot be collated in a single place for training a central model. All the devices that contributed towards training a global model have the right to use it for their own predictions. Google's GBoard uses personal keystrokes from multiple mobile phones to train a Next Word Prediction model via Federated Learning. BrainTorrent \cite{roy2019braintorrent} uses Federated Learning in the domain of medicine to perform the task of whole brain segmentation in a decentralized setting. \par

\subsection{Multiple Data Multiple Model}
This category encompasses solutions that leverage a globally learnt model via Federated Learning and further personalize it to suit different clients \cite{NEURIPS2020_f4f1f13c}. All participants own their own finetuned version of a global model for their individual use. Personalized Federated Learning can be achieved by treating the task like a meta-learning \cite{DBLP:journals/corr/abs-1909-12488} or multi-task learning \cite{NIPS2017_6211080f} problem or by mixing the global and local models. \par

\noindent
Examining the ownership and usage rights of data within cloud provider systems is crucial for ensuring robustness in design. Understanding the spread and statistics of the data allows for better management of access and usage policies, better anomaly detection during the model training process, etc. Implementing distributed training policies to leverage distributed data also needs to balance efficient communication strategies with effective learning strategies. This will help to create useful ML models that are amenable to a wider user base while still being technically robust.\par

\section{Taxonomy considering Defence Techniques}

\begin{figure}[b]
\centering
\includegraphics[scale=0.5]{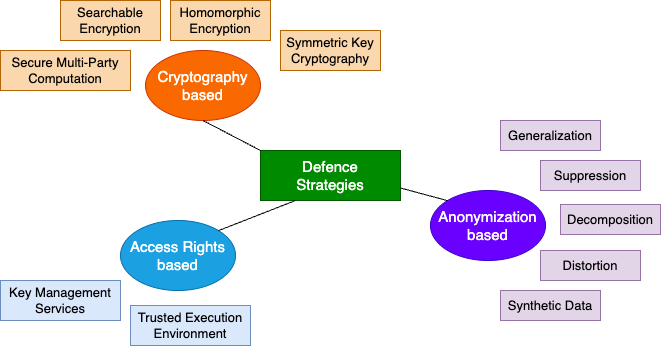}
\caption{Hierarchy of defences}
\label{fig:defence_taxonomy}
\end{figure}

\noindent
The need for different ML defences in the real world arises from the diverse range of security, privacy, fairness, and operational challenges that ML systems face. These defences are essential to ensure the responsible and secure deployment of ML models in critical applications and to maintain user trust and confidence in AI technologies. \par

A categorization of attack types that ML models face is presented in Table \ref{tab:all_attacks}. The defense strategies employed to mitigate these specific attacks can be classified into 3 buckets.
\begin{enumerate}
    \item \textbf{Cryptography based} defences guarantee confidentiality and integrity of data. 
    \begin{itemize}
        \item Symmetric key cryptography algorithms like AES, Blowfish etc. and asymmetric key cryptography algorithms like RSA, Diffie-Hellman etc. can be used to prevent unauthorized access to the training data and ML model. 
        \item Homomorphic Encryption performs computation on encrypted training data \cite{9407118, chakraborty2014functional} thus ensuring the confidentiality of the training data. Works like \cite{9734024} perform FHE on deep neural networks.
        \item Secure Multi Party Computation is used in privacy-preserving ML \cite{GAO201872, 10.1145/2810103.2813725} to prevent leakage of private information from distributed ML model training.
    \end{itemize}
 
    \item \textbf{Anonymization based} procedures mask personally identifiable information to avoid identity disclosure, thereby assuring confidentiality. Following are the main techniques available to implement anonymization.
    \begin{itemize}
        \item Generalization involves replacing specific or sensitive data values with more generalized or abstract ranges of data values. It obscures individual identities or characteristics while retaining statistical trends and patterns in the data.
        \item Suppression involves masking of specific data points or records from the dataset. This involves removing specific individuals with unique attributes or small groups that could be easily identified.
        \item Decomposition involves splitting a dataset into multiple parts or tables, each containing a subset of the original data. It can enhance privacy by separating sensitive attributes from less sensitive ones, making it more difficult to link them.
        \item Interference or Distortion introduces noise or random data into the dataset to mask sensitive information. It can be performed by adding random noise or multiplying by a constant to inflate actual values. This makes it more challenging to discern individual details or patterns, even when some data remains identifiable. Differential Privacy is one way to achieve ML privacy by using such data perturbation.
        \item Synthetic Data can be generated via statistical techniques using median or standard deviation, or classical linear regression. Synthetic image data can also be generated using Variational Auto Encoders (VAEs), Generative Adversarial Networks (GANs) etc. These can be shared in lieu of private and sensitive images for training ML models.
    \end{itemize}
 
    \item \textbf{Access Restrictions based} controls ensure that procedures perform authentication, authorization and maintain access control policies to regulate who has access to different parts of the system. Such enforcement can reduce the impact of risk. 
    \begin{itemize}
        \item Key Management Services provide secure environments for generating cryptographic keys, and also supports Customer managed keys via ‘Bring Your Own Key’ initiatives to ensure consistent key management policies.
        \item Trusted Execution Environments create isolated, secure regions within a computing device to provide a high level of security and privacy to perform sensitive tasks.
    \end{itemize}
\end{enumerate}

\begin{figure*}[b!]
\centering
  \includegraphics[scale=0.4]{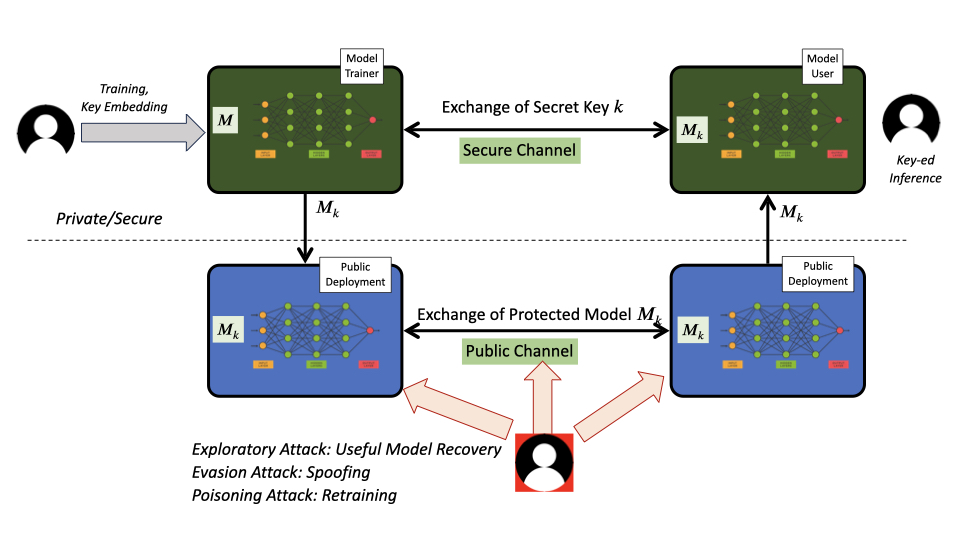}
  \caption{Encrypting an ML model.}
  \label{fig:cryptoML}
\end{figure*}

\begin{table*}[ht!]
\caption{Mapping known Privacy and Robustness attacks on Data and Model onto the proposed taxonomy of Section \ref{sec:data_model_taxonomy}.}
\resizebox{\columnwidth}{!}{%
\begin{tabular}{p{1.5cm}p{1cm}p{1cm}p{2cm}p{3cm}p{2.5cm}p{4cm}}
 \toprule
 \textbf{Setting} & \textbf{Data Owner} & \textbf{Model Owner} & \multicolumn{2}{c}{\textbf{Attacks}} & \multicolumn{2}{c}{\textbf{Defences}}\\
 \midrule
 \multirow{3}*{SDSM} & \multirow{3}*{1} & \multirow{3}*{1} & \multirow{3}*{Model - Privacy} & Model inversion & \multirow{2}*{Cryptographic} & Data-in-transit encryption \\
 & & & & Model reconstruction & & Compute on encrypted data \\
 & & & & Model extraction & Access rights based & Trusted Execution Environment \\
 \midrule
 
 \multirow{3}*{SDMM} & \multirow{3}*{1} & \multirow{3}*{n} & \multirow{3}*{Model - Privacy} & Model inversion & \multirow{2}*{Cryptographic} & Data-in-transit encryption \\
 & & & & Model reconstruction & & Compute on encrypted data \\
 & & & & Model extraction & Access rights based & Trusted Execution Environment \\
 \midrule
 
 \multirow{9}*{MDSM} & \multirow{9}*{n} & \multirow{9}*{1} & \multirow{3}*{Model - Privacy} & Model inversion & \multirow{2}*{Cryptographic} & Data-in-transit encryption \\
 & & & & Model reconstruction & & Compute on encrypted data \\
 & & & & Model extraction & Access rights based & Trusted Execution Environment \\
  & & & \multirow{5}*{Data - Privacy} & Infer class representatives & \multirow{2}*{Distortion} & \multirow{2}*{Differential Privacy} \\
  & & & & Infer statistical properties & & \\
  & & & & Infer attributes & \multirow{4}*{Cryptographic} & Compute on encrypted data \\
  & & & & Reconstruct training input and labels & & Symmetric Key cryptography \\
  & & & & Membership Inference & & Data-in-transit encryption \\
  & & & Data - Robustness & Data poisoning & & Symmetric Key cryptography \\
 \midrule
 
 \multirow{9}*{MDMM} & \multirow{9}*{n} & \multirow{9}*{m $\le$ n} & \multirow{3}*{Model - Privacy} & Model inversion & \multirow{2}*{Cryptographic} & Data-in-transit encryption \\
 & & & & Model reconstruction & & Compute on encrypted data \\
 & & & & Model extraction & Access rights based & Trusted Execution Environment \\
  & & & \multirow{5}*{Data - Privacy} & Infer class representatives & \multirow{2}*{Distortion} & \multirow{2}*{Differential Privacy} \\
  & & & & Infer statistical properties & & \\
  & & & & Infer attributes & \multirow{4}*{Cryptographic} & Compute on encrypted data \\
  & & & & Reconstruct training input and labels & & Symmetric Key cryptography \\
  & & & & Membership Inference & & Data-in-transit encryption \\
  & & & Data - Robustness & Data poisoning & & Symmetric Key cryptography \\
  & & & Model - Robust- & Model poisoning & \multirow{2}*{Access rights based} & \multirow{2}*{Trusted Execution Environment} \\
  & & & ness & Backdoor attacks & & \\
 \bottomrule
\end{tabular}
\label{tab:attacks_with_taxonomy}
}
\end{table*}

The user is afforded the flexibility to deploy one or more of the above-mentioned defence measures, facilitating a customized implementation aligned with their distinct security requirements. Figure \ref{fig:defence_taxonomy} arranges these measures hierarchically, according to the level at which the defence is required. Cryptography based defences are required at the data level, especially when the resources belonging to 1 user must be secured. Anonymization based defences apply well to groups of users, while Access rights based defences work at the application or API service level.\par

Figure \ref{fig:cryptoML} depicts a potential defence strategy that leverages cryptographic measures to transmit the private and protected ML model over public channels. This involves secret key exchange over secure channels, designed to embed the keys into the architecture of the neural network model without disrupting training layers. \par

\noindent
A comprehensive view of the proposed taxonomy along with recommended defence techniques is presented in Table \ref{tab:attacks_with_taxonomy} to foster a user-centric approach to ML solutioning. Consider the example of a membership inference attack on a mobile application used to predict the next word while users type on the device keyboard. Since the primary medium of information entry in mobile devices is the keyboard, it is privy to a lot of private and sensitive user information if all the keystrokes are recorded and used to train a centralized Next Word Prediction model. This environment context corresponds to the Multiple Data Single Model (MDSM) setting of the proposed taxonomy. Following Table \ref{tab:attacks_with_taxonomy}, membership inference attack is a type of attack on data privacy and the recommended defence for it can be either cryptographic or distortion based, by implementing Differential Privacy. \par
For model consumers of AIaaS, such a table provides a user-friendly reference guide that helps them identify security gaps in their solution architecture. They can implement practical solutions or guidance to resolve these issues independently or leverage out-of-the-box services offered by the cloud provider. On the other hand, model providers can benefit from this view by creating the building blocks which consumers will need to build robust and secure applications. It grants a competitive edge to the cloud providers who have a more extensive portfolio. \par

\section{Concluding Remarks}
\noindent 
As machine learning becomes increasingly pervasive in all domains, we realize the importance of educating the general public on the involved nature of privacy and security vulnerabilities in ML applications.
Recognizing the privacy risks of tools used today can also influence legislature and give more rights to the public. In a strong move towards the European Union's GDPR right to delete personal data, Amsterdam developed software that blurs people in images captured by their public street mapping cameras. Even though these images were used to train ML models, individuals' privacy is protected by concealing their face and skin colour. 
There is a pressing need to understand the evolving landscape of security threats and their defences in the context of AI enabled personal and critical applications.\par
We conducted a comprehensive study of the privacy guarantees offered by major cloud service providers and proposed a threat model to identify the systemic risks. We proposed a taxonomy to classify ML applications based on ownership of training data and usage rights of the trained model. Then we presented an extensive list of attacks possible to data and model privacy as well as robustness and proposed a classification for the defence techniques against it. Finally we unified the initial taxonomy with the different types of attacks and defences based on their characteristics.

\section*{Acknowledgement}
\noindent 
This research is supported by the National Research Foundation, Singapore under its Strategic Capability Research Centres Funding Initiative. Any opinions, findings and conclusions or recommendations expressed in this material are those of the author(s) and do not reflect the views of National Research Foundation, Singapore.

\bibliography{sigproc} 
\bibliographystyle{ieeetr}

\end{document}